# Vibrational resonance in narrow-band conductors

G.M. Shmelev[1]*, E.M. Epshtein[2], A.S. Matveev[3]

[1]Volgograd State Pedagogical University, Volgograd, Russia
[2]Institute of Radio Engineering and Electronics of the Russian Academy of Sciences, Fryazino, Russia
[3]Volgograd Branch of the Moscow Consumer Cooperation University

Abstract. The response is studied of a narrow-band conductor with bcc lattice to a low-frequency signal under presence of a high-frequency signal. In a high enough dc electric field $E_x$, the conduction electrons form a bistable system, which results in spontaneous appearance of a transverse electric field $E_y$. Ac field along $y$ axis leads to amplification of the low-frequency signal with non-monotonous dependence of the gain on the high-frequency field amplitude, so that vibrational resonance takes place. Besides, high-frequency field induced nonequilibrium phase transition is found.

Nonlinear phenomena in multistable systems are of great interest in the contemporary physics. Such phenomena conclude so-called stochastic resonance. Under simultaneous action of a weak harmonic signal and a noise upon a bistable system, the signal amplitude amplification is possible, and the gain as a function of the noise intensity has a maximum at some value of the latter [1 – 3]. Similar effect can be initiated by high-frequency vibration, instead of noise; that effect has been named as vibration resonance [4, 5]. Under such conditions, the hopping rate between stable states is determined by the vibration frequency, rather than noise, so that the resonance is more pronounced.

In present work, we consider vibrational resonance in a cubic crystal with a narrow conduction band, which is placed in a high enough dc electric field. Note that stochastic resonance in nonequilibrium electron gas of a quasi-two-dimensional semiconductor superlattice was considered for the first time in Refs. 6 and 7. In such materials, conduction electrons in a high electric field form a bistable system. It leads to a second type nonequilibrium phase transition (NPT) that manifests itself as spontaneous appearance of an electric field perpendicular to the current in the sample [8 – 10]. The phenomena considered in Refs. 6 to 10 are possible not only in quasi-two-dimensional superlattices, but also in bulk materials [8]. Such materials are presented, in particular, with cubic crystals, in which the transverse spontaneous electric field $E_y$ is the order parameter, while the driving field $E_x$ is the control parameter (X and Y axes being directed along the principal axes of a bcc lattice or tilted under 45° angle with respect to the principal axes of a simple cubic lattice). For definiteness, we consider the electron dispersion law in bcc lattice under tight-binding approximation:

$$\varepsilon(\mathbf{p}) = \varepsilon_0 - \Delta \cos\left(\frac{p_x d}{\hbar}\right)\cos\left(\frac{p_y d}{\hbar}\right)\cos\left(\frac{p_z d}{\hbar}\right), \quad (1)$$

where $\varepsilon(\mathbf{p})$ and p are electron energy and crystal momentum, respectively, $2\Delta$ is the conduction band width, $2d$ is the lattice constant.

In quasi-classical situation, $2\Delta \gg \hbar/\tau$, $|eE|d$ ($\tau$ is electron momentum relaxation time, e is the electron charge), the steady distribution function of the uniform electron gas, $f(p)$, can be obtained by means of the classical Boltzmann equation with the $\tau$-approximation for the collision integral:

$$\left(e\mathbf{E}, \frac{\partial f(\mathbf{p})}{\partial \mathbf{p}}\right) = \frac{f_0(\mathbf{p}) - f(\mathbf{p})}{\tau}, \quad \tau = const, \quad (2)$$

where $f_0(p)$ is the equilibrium distribution function.

---

* e-mail: shmelev@fizmat.vspu.ru

Note that refusal of the $\tau$-approximation does not lead to any principally new results for NPT. The solution of Eq. (2) takes the form

$$f(\mathbf{p}) = \int_0^\infty f_0(\mathbf{p} - e\mathbf{E}t) \exp(-t/\tau) dt/\tau. \quad (3)$$

The current density is

$$\mathbf{j} = e \sum_\mathbf{p} \frac{\partial \varepsilon(\mathbf{p})}{\partial \mathbf{p}} f(\mathbf{p}). \quad (4)$$

The calculation result can be presented by means of a synergetic potential [8] $\Phi(\mathbf{E}) = \int j_y dE_y + const$:

$$\mathbf{j} = \frac{\partial \Phi}{\partial \mathbf{E}}, \quad (5)$$

$$\Phi(\mathbf{E}) = \frac{\sigma_0 C_{111} E_0^2}{8} \times$$
$$\times \left[ \ln\left( \left(E_0^2 + E_x^2 + (E_y - E_z)^2\right)^2 - 4E_x^2(E_y - E_z)^2 \right) + \right.$$
$$+ \ln\left( \left(E_0^2 + E_x^2 + (E_y + E_z)^2\right)^2 - 4E_x^2(E_y + E_z)^2 \right) -$$
$$\left. - 2\ln(E_0^4) \right] + const \quad (6)$$

where $E_0 = \hbar/ed\tau$, $\sigma_0 = e^2 n \Delta d^2 \tau/\hbar^2$, $\sigma_0 C_{111}$ is low-field conductivity, $n$ is carrier density, $C_{111} = \left\langle \cos\left(\frac{p_x d}{\hbar}\right) \cos\left(\frac{p_y d}{\hbar}\right) \cos\left(\frac{p_z d}{\hbar}\right) \right\rangle$, the angular brackets mean averaging over the equilibrium distribution. In the case of the nondegenerate electron gas, $C_{111} \approx 1.025 [I_1(\Delta/2k_0 T)/I_0(\Delta/2k_0 T)]^2$, where $I_n(z)$ is the modified Bessel function, $k_0$ is the Boltzmann constant. Note that $\Phi(E_x, E_y, E_z)$ function is invariant with respect to permutations of $x, y, z$ indices.

First, consider a situation, when only dc electric field presents, **E** vector lies in $XY$ plane ($E_z = 0$), and the sample is open in $Y$ direction, so that

$$j_y = \frac{\partial \Phi}{\partial E_y} = 0. \quad (7)$$

It follows from Eqs. (5) – (7)

$$E_y \equiv E_{ys} = \begin{cases} 0, & 0 \leq |E_x| < E_0 \\ \pm\sqrt{E_x^2 - E_0^2}, & |E_x| \geq E_0 \end{cases}. \quad (8)$$

At $|E_x| \geq E_0$, the zero solution is unstable with respect to small fluctuations of the electric field. The nonzero solution in Eq. (8) describes the spontaneous electric field mentioned.

In terms of the synergetic potential, the stability condition takes the form [11]

$$\frac{\partial j_y}{\partial E_y} = \frac{\partial^2 \Phi}{\partial E_y^2} > 0. \quad (9)$$

At $|E_x| > E_0$, the potential $\Phi$ has two minima $E = E_{ys}$, which are stable, in accord with Eq. (9).

Near NPT point, i.e. within the critical range $|E_x^2 - E_0^2| \ll E_0^2$, the usual Landau expansion of the $\Phi$ potential up to terms $\sim E_y^4$ in the linear approximation on $|E_x^2 - E_0^2|$ takes the form

$$\Phi = \frac{(1 - E_x^2)}{(E_x^2 + 1)^2} \frac{E_y^2}{2} + \frac{1}{(E_x^2 + 1)^2} \frac{E_y^4}{4} + const(E_y). \quad (10)$$

Formula (10) is presented in a dimensionless form by means of the following substitutions: $\Phi/\sigma_0 C_{111} E_0^2 \to \Phi$, $E_{x,y}/E_0 \to E_{x,y}$.

Let two harmonic signals be applied to the sample along $Y$ direction. Then condition (7) is replaced by the following one:

$$\frac{dE_y}{dt} = -\frac{\partial \Phi}{\partial E_y} + A\cos(\omega t) + B\cos(\Omega t + \Theta), \quad (11)$$

where $A$ and $\omega$ are the low-frequency current amplitude and frequency, respectively, $B$, $\Omega$, and $\Theta$ are amplitude, frequency and phase, respectively, of the high-frequency current ($\Omega \gg \omega$). In Eq. (11), $4\pi\sigma_0 C_{111} t/\varepsilon \to t$ substitution is carried out ($\varepsilon$ being the dielectric constant), the frequencies are changed similarly, $A$ and $B$ amplitudes are expressed in $\sigma_0 E_0$ units.

We seek the solution of Eq. (11) in the following form (cf. [5]):

$$E_y = \widetilde{E}_y + \Psi(t, \Omega t), \qquad (12)$$

where the first term corresponds to slow change of $E_y$, while the second one is $2\pi$-periodical function of the "fast" time $\xi = \Omega t$, so that

$$\overline{\Psi(t, \Omega t)} = \frac{1}{2\pi} \int_0^{2\pi} \Psi(t, \xi) d\xi = 0. \qquad (13)$$

Under $d\Psi/dt \gg \Psi, \Psi^2, \Psi^3$ condition, substitution of Eq. (12) in Eq. (11) gives after some manipulation

$$\Psi = \frac{B}{\Omega} \cos(\Omega t + \Theta), \qquad (14)$$

$$\frac{d\widetilde{E}_y}{dt} + \frac{\partial U}{\partial \widetilde{E}_y} = A\cos(\omega t), \qquad (15)$$

where $U$ is an effective potential. In accordance with Eq. (10), the potential takes the form

$$U = -\frac{1}{(E_x^2+1)^2}(E_x^2 - E_{xb}^2)\frac{\widetilde{E}_y^2}{2} + \\ + \frac{1}{(E_x^2+1)^2}\frac{\widetilde{E}_y^4}{4} + \mathrm{const}(\widetilde{E}_y) \qquad (16)$$

$$E_{xb}^2 = 1 + \frac{3}{2}\left(\frac{B}{\Omega}\right)^2, \qquad (17)$$

where $E_{xb}$ is the renormalized bifurcation point. The potential minima correspond to the following values:

$$\widetilde{E}_y \equiv \widetilde{E}_{ys} = \begin{cases} 0, & E_x \le E_{xb} \\ \pm\sqrt{E_x^2 - E_{xb}^2}, & E_x > E_{xb} \end{cases}. \qquad (18)$$

The high-frequency field causes the bifurcation point shift by $|\Delta E_{xb}| = |E_{xb} - 1| \approx 3B^2/4\Omega^2 \ll 1$.

Note that an additional control parameter, $B/\Omega$, appears in consideration besides $E_x$. Indeed, $\widetilde{E}_y$ function with $B/\Omega = \mathrm{fix}$ displays behavior shown in Fig. 1, while its behavior at $E_x = \mathrm{fix}$ is shown in Fig. 2. The latter demonstrates the high-frequency current-induced second type NPT; the situation is similar to the noise-induced NPT [12].

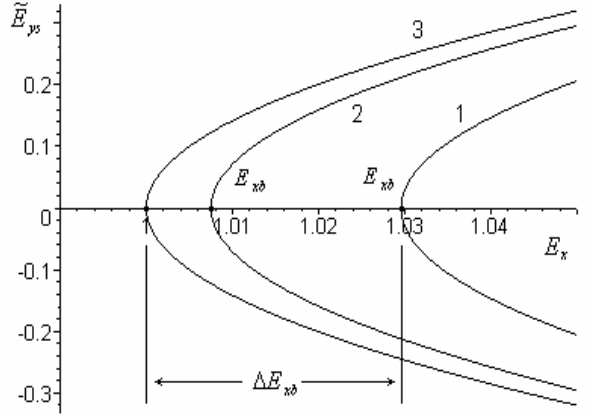

Fig. 1. Spontaneous transverse field as a function of the driving field (in per unit). 1 – $B/\Omega = 0.2$; 2 – $B/\Omega = 0.1$; 3 - $\widetilde{E}_{ys}(B=0) = E_{ys}$.

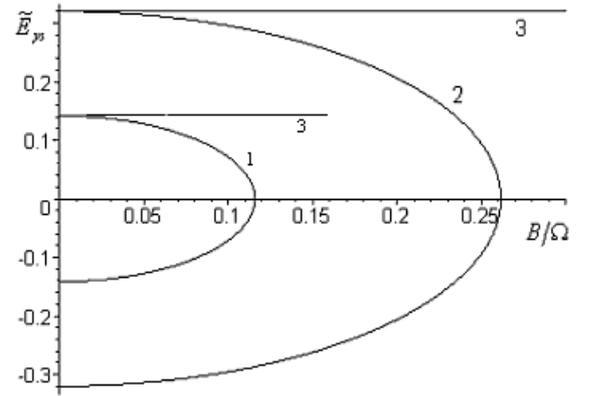

Fig. 2. Spontaneous transverse field as a function of the control parameter $B/\Omega$. 1 – $E_x = 1.01$; 2 – $E_x = 1.05$; 3 - $E_{ys} = \sqrt{E_x^2 - 1}$.

The low-frequency current in Eq. (15) causes forced oscillation of the $\widetilde{E}_y$ field with $\Delta\widetilde{E}_y = \widetilde{E}_y - \widetilde{E}_{ys}$ amplitude near the equilibrium value $\widetilde{E}_{ys}$. At $A \ll 1$ and $E_x > E_{xb}$, we obtain from Eq. (15) within linear approximation on $\Delta\widetilde{E}_y$

$$\frac{d(\Delta\widetilde{E}_y)}{dt} + 2\mu\Delta\widetilde{E}_y = A\cos(\omega t), \qquad (19)$$

$$\mu = \frac{(E_x^2 - E_{xb}^2)}{(E_x^2+1)^2}, \qquad (20)$$

where $(2\mu)^{-1}$ is dynamical relaxation time [3] near the stable state of the system. The stationary solution of Eq. (19) is

$$\Delta\widetilde{E}_y = \frac{A}{\sqrt{4\mu^2+\omega^2}}\cos(\omega t - \varphi), \quad \tan\varphi = \frac{\omega}{2\mu}. \quad (21)$$

The ratio of the low-frequency response amplitude $\Delta\widetilde{E}_y$ to the applied low-frequency signal amplitude is

$$Q = 1/\sqrt{4\mu^2+\omega^2} \quad (E_x > E_{xb}). \quad (22)$$

At $\omega = 0$, we have $Q = (E_x^2+1)^2/2(E_x^2 - E_{xb}^2)$. In the case of a single-well potential ($E_x < E_{xb}$), the Eq. (15) for $\Delta\widetilde{E}_y$ shift takes the form

$$\frac{d(\Delta\widetilde{E}_y)}{dt} + \mu\Delta\widetilde{E}_y = A\cos(\omega t), \quad (23)$$

while

$$Q = 1/\sqrt{\mu^2+\omega^2} \quad (E_x < E_{xb}). \quad (24)$$

Equations (22) and (23) with divergence at the bifurcation point with $\omega = 0$ agree completely with the corresponding conclusions of the second type phase transition Landau theory [13].

Typical behavior of $Q$ parameter as a function of the driving field is shown in Fig. 3.

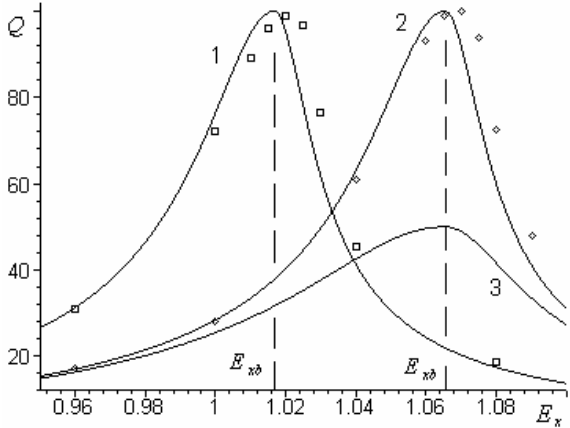

Fig. 3. Parameter $Q$ as a function of the driving field $E_x$: 1 – $B/\Omega = 0.15$, $\omega = 10^{-2}$; 2 – $B/\Omega = 0.3$, $\omega = 10^{-2}$; 3 – $B/\Omega = 0.3$, $\omega = 2\cdot 10^{-2}$. $\omega$ frequency is expressed in $4\pi\sigma_0 C_{111}/\varepsilon$ units. The squares and diamonds correspond to the numerical solution of Eq. (11) at $A=10^{-3}$.

The function maximum appears in the second type NPT bifurcation point at the field value $E_x = E_{xb}$. The vibrational resonance as a non-monotonous dependence of the gain on the field amplitude and frequency is shown in Figs. 4 and 5, respectively.

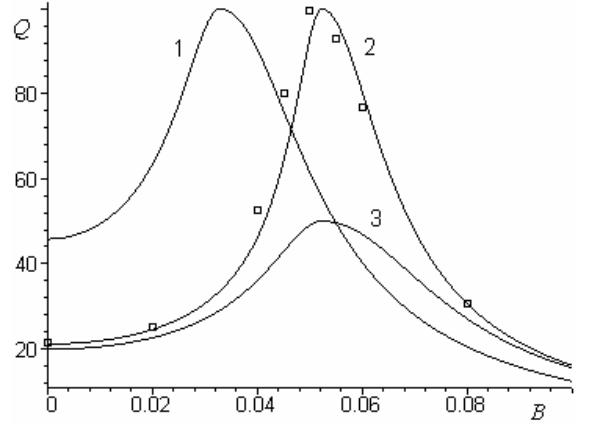

Fig. 4. Parameter $Q$ as a function of the high-frequency signal amplitude $B$. 1 – $E_x = 1.02$, $\Omega = 0.2$, $\omega = 10^{-2}$; 2 – $E_x = 1.05$, $\Omega = 0.2$, $\omega = 10^{-2}$; 3 – $E_x = 1.05$, $\Omega = 0.2$, $\omega = 2\cdot10^{-2}$. $\omega$ and $\Omega$ frequencies are expressed in $4\pi\sigma_0 C_{111}/\varepsilon$ units. The squares correspond to the numerical solution of Eq. (11) at $A=10^{-3}$.

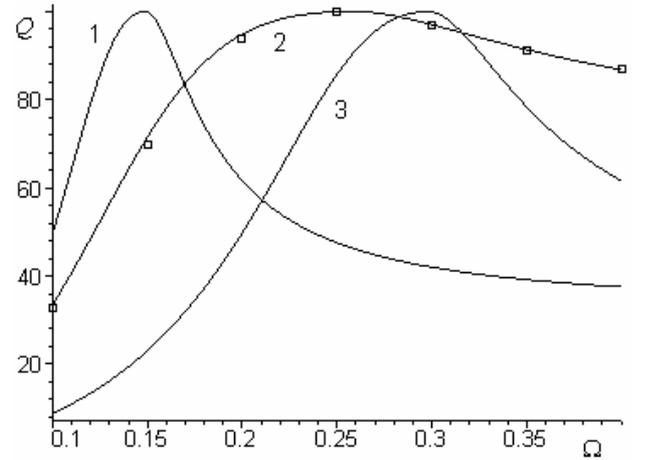

Fig. 5. Parameter $Q$ as a function of the high-frequency signal frequency $\Omega$. 1 – $E_x = 1.03$, $\omega = 10^{-2}$, $B = 0.03$; 2 – $E_x = 1.01$, $\omega = 10^{-2}$, $B = 0.03$; 3 – $E_x = 1.03$, $\omega = 10^{-2}$, $B = 0.06$. $\omega$ and $\Omega$ frequencies are expressed in $4\pi\sigma_0 C_{111}/\varepsilon$ units. The squares correspond to the numerical solution of Eq. (11) at $A=10^{-3}$.

The maxima correspond to the solution of Eq. (17). Note that the gain is more than 1 in absence of the high-frequency field (see Fig. 4), because a weak external periodical current is amplified at the expense of the energy of dc electric field $E_x$. In absence of the driving field ($E_x = 0$), the output signal amplitude is less than input one.

Let us make some estimation. At $d = 10^{-7}$ cm, $\tau = 10^{-12}$ s, $\varepsilon = 10$, $\Delta = 10^{-2}$ eV, $T = 300\ K$, $n = 10^{16}$ cm$^{-3}$, we have $E_0 \approx 6600$ V/cm, $\varepsilon/(4\pi\sigma_0 C_{111}) \approx 2.3\times 10^{-10}$ s. Note that $E_0$ is smaller by an order of magnitude in artificial crystals (three-dimensional superlattices) with $d = 10^{-6}$ cm.

The work was supported by RFBR grant No. 02-02-16238.